\documentclass[a4paper]{jpconf}
\usepackage{graphicx}
\begin{document}
\title{Mesons, PANDA and the scalar glueball}

\author{Denis Parganlija}

\address{Vienna University of Technology, Institute for Theoretical Physics\\Wiedner Hauptstr.\ 8-10/E136, 1040 Vienna, Austria}

\ead{denisp@hep.itp.tuwien.ac.at}

\begin{abstract}
The non-perturbative nature of Quantum Chromodynamics (QCD) at low energies has prompted the expectation that the gauge-bosons of QCD -- gluons --
might give rise to compound objects denoted as glueballs. Experimental signals for glueballs have represented a matter of research for various
collaborations in the last decades; future research in this direction is a main endeavour planned by the PANDA Collaboration at FAIR. Hence in this article I review
some of the outstanding issues in the glueball search, particularly with regard to the ground state -- the scalar glueball, and discuss the relevance for PANDA at FAIR.
\end{abstract}

\section{Introduction}
Approximately five decades ago a truly remarkable number of new, strongly interacting resonances was discovered leading way to a new direction in the development
of nuclear physics (a non-exhaustive list of the relevant articles is presented in Ref.\ \cite{6d}
and further work can be traced in the Particle Data Tables \cite{PDG}). These discoveries
coincided with the appearance of many renowned articles in theoretical physics -- a selection of which can be found in Ref.\ \cite{theory} --
that represented a basis for the development of a theory of strong interactions. The most viable approach was subsequently found to be the one where
``the strong interactions are described by an unbroken gauge theory based on
tricolored quarks and color octet vector gluons, with color supposed to be entirely
confined, so that quarks are fractionally charged'' with the suggestion that ``a good name for this theory
is quantum chromodynamics (QCD)'' \cite{Fritzsch:1975sr}.\\
\\
Quantum Chromodynamics has various interesting features, one of the most important of which is a scale-dependent coupling \cite{AF} of such nature that, at higher energies,
it leads to weaker interactions between the relevant degrees of freedom -- quarks and gluons. Conversely, at lower energies, quarks and gluons are confined into
more macroscopic structures: hadrons. According to the total spin $J$, hadrons are divided into mesons (that possess integer spin) and baryons (that possess half-integer spin). Determination of
hadron properties from Quantum Chromodynamics is a highly non-trivial task due to the inapplicability of perturbative approaches in the energy region where hadrons emerge.
A circumvention of this problem is introduced by effective models containing the so-called ``constituent'' quarks and gluons that, in the particular case of mesons, allow for the particle 
classification as quarkonia ($\bar{q}q$, see Ref.\ \cite{PhD} and references therein), tetraquarks ($\bar{q}\bar{q} qq$ \cite{Jaffeq2q2}), hybrids 
(constituent quarks with excited gluon degrees of freedom \cite{Barnes:1982tx}) and others.
\\
%\\
The non-Abelian nature of QCD implies gluon self-interaction that is governed by the same scale-dependent coupling as in the
quark sector. Hence the natural expectation is that, analogously to quarks, the gluons also bind into composite objects -- the glueballs \cite{glueballs}.
\\
\\
There are at least two important reasons to consider glueballs as low-energy degrees of freedom in QCD:
\begin{itemize}
\item Glueball mass is generated exclusively by the gluon self-interaction. The Higgs
mechanism does not contribute to the mass generation of glueballs rendering them uniquely suitable
for the research of the strong interaction.

\item Gluons are vector particles and hence glueballs possess integer spin. This implies that glueballs belong
to the meson spectrum -- and therefore the meson spectrum can only be complete if the spectrum of glueballs is known.

\end{itemize}
As already noted, glueballs have long been subject to experimental and theoretical research.
%\section{Glueballs, experiment and theory}
In general, low-energy mesons are mainly produced in these channels \cite{PhD,Hersonissos}:
\begin{itemize}

\item antiproton-proton ($\bar{p}$-$p$) collisions \cite{barp-p}

\item proton-proton ($p$-$p$) collisions \cite{p-p,WA76}

\item electron-positron annihilation into $J/\psi$ or $\varphi(1020)$ and their subsequent decay into further meson
resonances \cite{e+e-}

\item pion-nucleon scattering reactions \cite{WA76,pi-N}.

\end{itemize}
A glueball state is expected to be preferably produced in radiative decays and to possess a strongly suppressed $\gamma \gamma$
decay (or a strongly suppressed production in the $\gamma \gamma$ collisions).\\
\\
On the theory side, there are various means suitable for the glueball identification. These range from
effective models \cite{glueballs,models} to
lattice QCD \cite{lattice} and holographic-QCD approaches \cite{holography}.
\\
\\
However, a clear experimental signal for a glueball can only be obtained if the resonance has a strongly suppressed mixing to other states with 
the same quantum numbers.
%This implies, for example, that the so-called exotic mesons -- i.e., those with quantum numbers inaccessible
%to $\bar{q}q$ configurations -- suffer only marginally from mixing effects rendering relatively clear signals.
Otherwise the glueball search can become massively complicated, as we illustrate in the following on the example of the scalar glueball.

\section{The scalar glueball}

Scalar mesons are QCD degrees of freedom that possess quantum numbers $J^{PC} = 0^{++}$,
where $J$ is the total spin, $P$ represents parity and $C$ is the charge 
conjugation; mesons with isospin $I=0$ are denoted as isoscalars.
\\
PDG data suggest the existence of 
five $IJ^{PC} = 00^{++}$ states in the energy region up to 1.8 GeV: $f_0(500)$ or $\sigma$, $f_0(980)$, $f_0(1370)$, $f_0(1500)$ and $f_0(1710)$ \cite{PDG}.
At the level of quantum numbers, all the mentioned states are viable candidates for the scalar glueball. %The expectation is that
Then the glueball
should be identified by analysing ({\it i}) the production mechanism and ({\it ii}) the decay channels.\\
\\
For the production mechanism, the situation is as follows \cite{PDG}:
\begin{itemize}
 \item Produced in radiative decays: fulfilled by $f_0(980)$, $f_0(1500)$ and $f_0(1710)$.
 
 \item Not seen in $\gamma \gamma$ collisions: fulfilled by $f_0(1500)$ and $f_0(1710)$.
 
\end{itemize}
Then $f_0(1500)$ and $f_0(1710)$ appear to be the strongest candidates for the scalar glueball since they fulfill both of the above criteria.
At this point, let us briefly comment on the above $\gamma \gamma$ data since they are of great importance for the glueball search and also
since they allow us to illustrate experimental ambiguities in the $f_0$ sector. The main reason to include
$f_0(1500)$ and $f_0(1710)$ into the above list is the fact that there was a notable absence of both of these states in the $\gamma\gamma$ data analysed by the ALEPH
Collaboration in 2000 \cite{ALEPH}.
Contrarily, an independent search by the L3 Collaboration -- that approximately coincided with the ALEPH results -- observed a clear signal
for $f_0(1710)$ while no signal was found for $f_0(1500)$ \cite{L3}. These are currently the best, and the latest, experimental data containing
negative results on photon coupling of the $f_0$ states, but they do not appear to be conclusive.
\\
\\
Hence the expectation might be that the decay channels of the $f_0$ resonances should provide us with more clear-cut results for the glueball identification.
The reason is that, in the case of the unbroken flavour symmetry, one can calculate exact decay-width ratios valid for a glueball state without mixing
to states containing quarks (``pure  glueball''). Then, for example, the ratio of the pion decay width to the kaon decay width is 3/4.\\
Unfortunately, neither the data on $f_0(1500)$ nor those on $f_0(1710)$ follow this expectation:
\begin{itemize}
 \item The $f_0(1500)$ resonance decays predominantly into pions ($2\pi$ and $4\pi$) whereas the kaon decay is suppressed;
the ratio $\Gamma_{f_0(1500)\rightarrow \pi\pi} / \Gamma_{f_0(1500)\rightarrow KK} \simeq 4$ \cite{PDG}.

\item The $f_0(1710)$ resonance decays predominantly into kaons and the pion decay is suppressed; the ratio
$\Gamma_{f_0(1710)\rightarrow \pi\pi} / \Gamma_{f_0(1710)\rightarrow KK} \simeq (0.2 - 0.4)$, depending on collaboration \cite{PhD}.

\end{itemize}
These results demonstrate clearly that there is no exclusive glueball state in the scalar sector 
-- one can only identify a resonance
that is \emph{predominantly} a glueball. Then, as already indicated, various model calculations have to be utilised in order to study mixing patterns of the pure glueball
with $\bar{q}q$, $\bar{q}\bar{q}qq$ and other states \cite{glueballs,models}. Lattice results \cite{lattice} are also of extreme importance, especially since the inclusion
of dynamical quarks has become more prominent; realistic approaches in holographic QCD can be very helpful to the cause as well \cite{holography}.
\\
\\
However, there is a further complication: the possible existence of a new scalar state, $f_0(1790)$.

\section{The \boldmath $f_0(1790)$ resonance and PANDA}

The BES II Collaboration claimed the existence of the new $f_0(1790)$ resonance in 2004 \cite{BESII2004}. The mass and the decay width of the 
resonance were determined by the Collaboration to be $m_{f_{0}(1790)}=1790_{-30}^{+40}$ and $\Gamma_{f_{0}(1790)}=270_{-30}^{+60}$ MeV, respectively.
The stated large decay width implies a strong overlap
with $f_0(1710)$. However, there is a clear point of distinction between the two resonances: as already mentioned, $f_0(1710)$ is reconstructed predominantly in the kaon
decay channels whereas $f_0(1790)$ is reconstructed predominantly in the pion decay channels.\\\\
The distinction between $f_0(1710)$ and $f_0(1790)$ is further ascertained by their production mechanism that involves $J/\psi$ decays \cite{BESII2004}:
({\it i}) $J/\psi\rightarrow\varphi K^+ K^-$;
({\it ii}) $J/\psi\rightarrow\varphi \pi^{+}\pi^{-}$;
({\it iii}) $J/\psi\rightarrow\omega K^{+}K^{-}$;
({\it iv}) $J/\psi\rightarrow\omega \pi^{+}\pi^{-}$.
The $f_0(1710)$ resonance is reconstructed from the decays ({\it i}) and ({\it iii})
whereas the decays ({\it ii}) and ({\it iv}) allow for the reconstruction of $f_{0}(1790)$.
Then assuming $f_{0}(1710)$ and $f_{0}(1790)$ to be the same resonance leads to a contradiction: 
such a resonance would have to possess a pion-to-kaon-decay ratio of $1.82 \pm 0.33$ according to reactions
({\it i}) and ({\it ii}) and, simultaneously, the pion-to-kaon-decay ratio $< 0.11$ according to reactions
({\it iii}) and ({\it iv}) \cite{BESII2004}. That is obviously not possible -- and hence
the BES II data suggest $f_{0}(1710)$ and $f_{0}(1790)$ to represent two distinct resonances.
\\
\\
Let us now discuss the importance of this statements for the PANDA experiments at FAIR. The PANDA Collaboration intends to
study interactions between antiprotons and fixed-target protons and nuclei in the momentum range of 1.5-15 GeV
with the stated goal of
exploring the spectroscopy of glueballs, multiquark and exotic states \cite{PANDA}.
\\\\
The search for glueballs is an extremely important endeavour of the Collaboration since, as already indicated, these states offer a unique 
insight into the strong interaction. However, there are numerous complications along this path, exhibited for example in the case of the ground state
-- the scalar glueball.
\\
This state will inevitably have strong mixing with mesons of $\bar{q}q$, $\bar{q}\bar{q}qq$ and other structures.
Nonetheless, it is extremely important to identify the scalar glueball since otherwise the glueball spectrum would not be complete.
In this case, however, one must bear in mind that
the most basic condition for the identification is to ascertain
\emph{the number} of the scalar states.
In the light of the BES II data, it is then of extreme importance for the PANDA Collaboration
to search for, and confirm -- or disprove -- the existence of, the $f_0(1790)$ resonance since, if this resonance exists, it will inevitably overlap
with the scalar glueball.

\section{Conclusions}
Glueball search is one of the most interesting challenges for the future PANDA experiments. These particles are expected to
be generated by the strong interaction and therefore the determination of their spectrum is of extreme importance for our understanding
of low-energy QCD. The determination of the ground, scalar, state is imperative for the completeness of the glueball spectrum.
Due to the mixing of the pure glueball state with other mesons that have the same quantum numbers, the scalar glueball can only be 
reliably identified if all scalar mesons ($f_0$ resonances) are known. At present, it is not clear whether this is the case: the existence of a new
resonance denoted as $f_0(1790)$ has been suggested by the BES II data
locating $f_0(1790)$ exactly in the energy region where the scalar glueball is expected.
Hence it is very important for the existence of
$f_0(1790)$ to be confirmed or disproved -- and the 
PANDA Collaboration is in a unique position to perform that task.

\ack
I am very grateful to D.\ Bugg, F.\ Giacosa, S.\ Janowski, P.\ Kovacs, A.\ Rebhan, D.\ H.\ Rischke and Gy.\ Wolf for extensive discussions
regarding meson spectroscopy.

\end{document}